\documentclass[aps,prb,12pt,tightenlines,%
notitlepage,longbibliography]{revtex4-1}
\usepackage{amsthm}
\usepackage{amsfonts}
\usepackage{hyperref}
\usepackage{graphicx}
\usepackage{color}
\newcommand\css{\mathop{\mathtt{CSS}}}
\begin{document}
\title%
%\chapter%[Topological defects in quantum LDPC codes]%
{Topological defects in general quantum LDPC codes}
\author%[Pak  Kau Lim]
{Pak Kau Lim}
%\aindx{Lim, Pak Kau}
\address{Department of Physics \&
      Astronomy,\\ University of California, Riverside,
    California, 92521 USA}

\author%[K. Shtengel]
{Kirill Shtengel}
\address{Department of Physics \&
      Astronomy,\\ University of California, Riverside,
    California, 92521 USA}
%\aindx{Shtengel, K.}
\author%[L. P. Pryadko]
{\underline{Leonid P. Pryadko}}
\address{Department of Physics \&
      Astronomy,\\ University of California, Riverside,
    California, 92521 USA}

%\aindx{Pryadko, L. P.}                          % author index entry
\begin{abstract}
  We consider the structure of defects carrying quantum information in
  general quantum low-density parity-check (LDPC) codes.  These
  generalize the corresponding constructions for topological quantum
  codes, without the need for locality.  Relation of such defects to
  (generalized) topological entanglement entropy is also discussed.
\end{abstract}
\maketitle
%\body
\def\ket#1{\left|#1\right\rangle}
\def\bra#1{\left\langle#1\right|}
\def\tr{\mathop{\rm tr}}
\def\sh{\mathop{\rm Sh}\nolimits}
\def\wgt{\mathop{\rm wgt}\nolimits}
\def\rank{\mathop{\rm rank}\nolimits}
\def\supp{\mathop{\rm supp}\nolimits}
\def\openone{{\bf 1}}

\newtheorem{theorem}{Theorem}
\newtheorem{statement}[theorem]{Statement}
\newtheorem{construction}{Construction}

\section{Introduction}
One of the many advantages of surface codes is the flexibility they
offer in the code parameters and the structure of logical operators.
To add an extra qubit one may simply create a hole in the surface.  A
larger hole, well separated from other defects, offers better
protection (larger minimal code distance).  Pairs of such holes can be
moved around to perform encoded Clifford gates,
etc\cite{Dennis-Kitaev-Landahl-Preskill-2002,%
  Bombin-MartinDelgado-2009,Bombin-2010}.

On the other hand, a substantial disadvantage of surface codes, or any
stabilizer code with generators local on a $D$-dimensional Euclidean
lattice, is that such codes necessarily have small rates $R=k/n$
whenever the code distance $d$ gets
large\cite{Bravyi-Terhal-2009,Bravyi-Poulin-Terhal-2010}.  Here $k$ is
the number of encoded qubits and $n$ is the block length of the code.
To get a finite asymptotic rate, one needs more general quantum codes.
In particular, any family of \emph{$w$-bounded quantum LDPC codes}
with stabilizer generators of weight not exceeding $w>0$ and distances
divergent as a logarithm or a power of $n$ has a non-zero asymptotic
error correction threshold even in the presence of measurement
errors\cite{Kovalev-Pryadko-FT-2013,Dumer-Kovalev-Pryadko-bnd-2015}.
Several families of bounded-weight quantum LDPC codes with finite
rates have been constructed.  Best-known constructions are quantum
hypergraph-product (qHP) and related codes\cite{Tillich-Zemor-2009,%
  kovalev-pryadko-hyperbicycle-2013,Zeng-Pryadko-2018}, and various
hyperbolic codes\cite{Zemor-2009,Delfosse-2013,%
  Guth-Lubotzky-2014,Breuckmann-Terhal-2015}.

The biggest obstacle to practical use of finite-rate quantum LDPC
codes is that their stabilizer generators must include far separated
qubits, regardless of the qubit layout in a $D$-dimensional
space\cite{Bravyi-Terhal-2009,Bravyi-Poulin-Terhal-2010}.  
Error correction requires frequent measurement of all stabilizer
generators, and measuring such non-local generators just isn't
practical if the hardware only allows local measurements.
Nevertheless, there is a question of whether other advantages of
surface codes, e.g., the ability to perform protected Clifford gates
by code deformations, can be extended to more general quantum LDPC
codes.

Such a construction generalizing the surface-code defects and gates by
code deformations to the family of qHP codes\cite{Tillich-Zemor-2009}
has been recently proposed by Krishna and
Poulin\cite{Krishna-Poulin-2019}.  However, their defect construction
is very specific to qHP codes.  Second, Krishna and Poulin do not
discuss the distance of the defect codes they construct, even though
it is important for the accuracy of the resulting gates.  Indeed,
since gates by code deformation are relatively slow, the distance has
to be large enough to suppress logical errors.

The purpose of this work is to give a general defect construction
applicable to any stabilizer code.  In the simplest form, one may just
remove a stabilizer generator which produces an additional logical
qubit, $k\to k+1$.  However, the distance $d'$ of such a code will not
exceed the maximum stabilizer generator weight, $d'\le w$.  Given a
degenerate quantum LDPC code with the stabilizer generator weights
bounded by $w$ and a distance $d>w$, we would actually like to
construct a related code encoding more qubits but retaining
degeneracy, i.e., with a distance $d'>w$. 
We propose a three-step defect construction: remove qubits in an
erasable region to obtain a subsystem code, do gauge-fixing to obtain
a stabilizer code with some generators of weight exceeding $w$, and
promote one or more such generators of the resulting code to logical
operators.  The choice of the gauge-fixing prescription is easier in
the case of Calderbank, Shor, and Steane (CSS)
codes\cite{Calderbank-Shor-1996,Steane-1996}, which makes the
construction more explicit.  For such codes, with some additional
assumptions, we give a lower bound on the distance of the defect code.
This shows that defect codes with unbounded distances can be
constructed, as is also the case with surface codes.

An interesting and a rather unexpected application of this analysis is
the relation of qubit-carrying capacity of a defect to its
(generalized) topological entanglement
entropy\cite{Levin-Wen-2006,Kitaev-Preskill-2006,%
  Grover-Turner-Vishwanath-2011} (TEE), denoted $\gamma$.  Namely, a
degenerate defect code with distance $d'>w$ can only be created when
$\gamma>0$.  Further, when distance $d'$ is large, the TEE $\gamma$
acquires stability: it remains non-zero whenever the defect is
deformed within certain bounds.

\section{Defect construction}
\label{sec:defect}
Generally, an $n$-qubit quantum code is a subspace of the $n$-qubit
Hilbert space $\mathbb{H}_{2}^{\otimes n}$.  A quantum $[[n,k,d]]$
stabilizer code is a $2^{k}$-dimensional subspace
$\mathcal{Q}\subseteq\mathbb{H}_2^{\otimes n}$ specified as a common
$+1$ eigenspace of all operators in an Abelian \emph{stabilizer} group
${\cal S}\in{\cal P}_{n}$, $-\openone\not\in{\cal S}$, where
$\mathcal{P}_n$ denotes the $n$-qubit Pauli group generated by tensor
products of single-qubit Pauli operators.  The stabilizer is typically
specified in terms of its generators,
${\cal S}=\left\langle S_{1},\ldots,S_{r}\right\rangle $.  If the
number of independent generators is $r\equiv\rank {\cal S}$, the code
encodes $k=n-r$ qubits.  The weight of a Pauli operator is the number
of qubits that it affects.  The distance $d$ of a quantum code is the
minimum weight of a Pauli operator $L\in{\cal P}_n$ which commutes
with all operators from the stabilizer ${\cal S}$, but is not a part
of the stabilizer, $L\not\in{\cal S}$. Such operators act
non-trivially in the code and are called logical operators.

 An $n$-qubit CSS stabilizer code
$\mathcal{Q}\equiv \css(P,Q)$ is specified in terms of two $n$-column
binary stabilizer generator matrices $H_X\equiv P$ and $H_Z\equiv Q$.
Rows of the matrices correspond to stabilizer generators of $X$- and
$Z$-type, respectively, and the orthogonality condition $PQ^T=0$ is
required to ensure commutativity.  The code encodes
$k=n-\rank P-\rank Q$ qubits, and has the distance $d=\min(d_X,d_Z)$,
\begin{equation}\label{eq:CSS-distance}
d_X=\min_{b\in \mathcal{C}_Q^\perp\setminus \mathcal{C}_P}\wgt(b),\quad 
d_Z=\min_{c\in \mathcal{C}_P^\perp\setminus  \mathcal{C}_Q}\wgt(c).
\end{equation}
Here $\mathcal{C}_Q\in\mathbb{F}_2^{\otimes n}$ is the binary linear
code (linear space) generated by the rows of $Q$, and
$\mathcal{C}_Q^{\perp}$ is the corresponding dual code formed by all
vectors in $\mathbb{F}_2^{\otimes n}$ orthogonal to the rows of $Q$.
Matrix ${Q}$ is the parity check matrix of the code
$\mathcal{C}_Q^\perp$.  A generating matrix of $\mathcal{C}_Q^\perp$,
${Q}^*$, has $\rank {Q}^*=n-\rank Q$ and is called \emph{dual} to $Q$.
Also, if $V=\{1,\ldots,n\}$ is the set of indices and $B\subset V$ its
subset, for any vector $b\in\mathbb{F}_2^{\otimes n}$, we denote
$b[B]$ the corresponding \emph{punctured} vector with positions
outside $B$ dropped.  Similarly, $Q[B]$ (with columns outside of $B$
dropped) generates the code $\mathcal{C}_Q$ \emph{punctured} to $B$.
We will also use the notion of a binary code $\mathcal{C}$
\emph{shortened} to $B$, which is formed by puncturing only vectors in
$\mathcal{C}$ supported inside $B$,
$$\text{Code $\mathcal{C}$ shortened to $B$ =} \{c[B]: \;c\in
\mathcal{C}\; \wedge\;\supp(c)\in B\}.$$ We will denote ${Q}_B$ a
generating matrix of the code ${\cal C}_Q$ shortened to $B$.  If $G$
and $H=G^*$ is a pair of mutually dual binary matrices, i.e., $GH^T=0$
and $\rank G+\rank H=n$, then $H_B$ is a parity check matrix of the
punctured code $\mathcal{C}_{G[B]}$, and\cite{MS-book}
\begin{equation}
  \label{eq:puncture-shortening-rank}
  \rank G[B]+\rank H_B=|B|.
\end{equation}
The distance $d$ of a linear code $\mathcal{C}$ is the minimal Hamming
weight of a non-zero vector in $\mathcal{C}$.  In general puncturing
reduces the code distance.  More precisely, if $d$ and $d'$ are the
distances of the original and the punctured code, respectively, they
satisfy $d-|A|\le d'\le d$.  On the other hand, the minimum distance
$d''$ of a shortened code is not smaller than that of the original
code, $d''\ge d$.

For a quantum code, if $A$ is a set of qubits and $B=V\setminus A$ its
complement, the stabilizer group $\mathcal{S}$ can also be punctured
to $B$, by dropping all positions outside $B$.  With the exception of
certain special cases\cite{Rains-1999,Sarvepalli-thesis-2008}, the
resulting group $\mathcal{G}\equiv \mathcal{S}[B]$ will not be
Abelian, and can be viewed as a gauge group of a subsystem
code\cite{Poulin-subs-2005,Bacon-subs-2006} called the \emph{erasure}
code.  A stabilizer code can be obtained by removing some of the
generators from $\mathcal{G}$ to make it Abelian; such a procedure is
called gauge-fixing.  In the case of a CSS code with stabilizer
generator matrices $H_X=P$ and $H_Z=Q$, the punctured group has
generators $P[B]$ and $Q[B]$, while a gauge-fixed stabilizer code can
be obtained, e.g., by replacing punctured matrix $Q[B]$ with the
corresponding shortened matrix, $Q_B$.  This latter construction can
be viewed as a result of measuring qubits outside $B$ in the $X$
basis.  Qubits in an erasable set $A$ can be removed without
destroying quantum information.  In this case, according to the
cleaning Lemma\cite{Bravyi-Poulin-Terhal-2010}, the logical operators
of the original code can all be chosen with the support outside $A$.
From here, with the help of Eqs.~(\ref{eq:CSS-distance}) and
(\ref{eq:puncture-shortening-rank}), one obtains (see Appendix
\ref{sec:proofs} for all proofs): %
\begin{statement}
  \label{th:CSS-decomposition}
  Consider a CSS code $\mathcal{Q}\equiv \css(P,Q)$ on qubit  set
  $V$ of cardinality $|V|=n$, 
  encoding $k$ qubits and with the CSS distances $d_X$, $d_Z$.  Let
  $A\subset V$ be an erasable in $\mathcal{Q}$ set of qubits, and
  $B\equiv V\setminus A$ its complement.  Then, the length-$|B|$ code
  $\mathcal{Q}'\equiv \css(P[B],Q_B)$ encodes the same number
  of qubits, $k'=k$, and has the CSS distances $d_X'$, $d_Z'$ such
  that:
  \begin{equation}
    d_X-|A|\le    d_X'\le d_X,\quad
    d_Z'\ge d_Z.\label{eq:distances}    
  \end{equation}
\end{statement}
The statement about the number of encoded qubits is true in general:
an erasure code and any of the corresponding gauge-fixed codes encode
the same number of qubits as the original code as long as the set $A$
of removed qubits is erasable.  (And, of course, we want to stick to
erasable sets since we do not want to lose quantum information).  To
construct a code that encodes $k''>k$ qubits, it is not sufficient to
just remove some qubits, one has to also remove some group generators.
If we do not care about the weight of stabilizer
generators and start with a generic stabilizer code, a code with a
decent distance may be obtained simply by dropping one of the existing
stabilizer generators.  Our general construction below is focused on
quantum LDPC codes with weight-limited stabilizer generators:

\begin{construction}
  \label{constr:defect} 
  Given an original $[[n,k,d]]$ degenerate code with stabilizer
  generator weights bounded by some $w<d$, in order to create a
  degenerate ``defect'' code with $k'>k$ and $d'>w$, {\rm ({\bf i})}
  remove some qubits in an erasable set, {\rm(\textbf{ii})} gauge
  fix the resulting subsystem code, and then {\rm(\textbf{iii})}
  drop one or more stabilizer generators with weights bigger than $w$.
\end{construction}
The gauge group $\mathcal{G}=\mathcal{S}[B]$ of the erasure code in
step (\textbf{i}) has generators of weights $w$ or smaller; generators
of weight greater than $w$ are obtained after gauge fixing in step
(\textbf{ii}).  This construction does not guarantee whether we get a
degenerate code or not.  Below, with the help of some additional
assumptions, we prove several inequalities that guarantee the
existence of not only degenerate defect codes with $d'>w$, but also
highly-degenerate defect codes with unbounded distances.  

\section{Distance bounds for a defect in a CSS code}
First, let us get general expressions for the distances $d_X'$, $d_Z'$
of a CSS code with a removed $Z$-type generator.  Given the original
code $\css(P,Q)$, we choose a linearly-independent row of $Q$,
${u}_0$, as the additional type-$Z$ logical operator, and denote
${Q}'$ the corresponding matrix with the row dropped (and of the rank
reduced by one).  Denote
\begin{equation}
d_Z^{(0)}=\min_\alpha \wgt(u_0+\alpha Q'),\label{eq:row-distance-Z}
\end{equation}
the minimum weight of a linear combination of $u_0$ with the 
rows of $Q'$.  Then, Eq.~(\ref{eq:CSS-distance}) gives 
\begin{equation}
  \label{eq:CSS-distance-Z}
  d_Z'=\min(d_Z,d_{Z}^{(0)}).
\end{equation}
The additional type-$X$ logical operator has to be taken from the set
of detectable errors of the original code.  Specifically, it has to
anticommute with the element of the stabilizer being removed, but
commute with the remaining operators in the stabilizer and all logical
operators of the original code.  In addition to the $X$-type logical
operators of the original code, the logical operators of the new code
include all errors with the same syndrome as the chosen canonical
operator.  Respectively, the expression for the distance reads:
\begin{equation}
  \label{eq:CSS-distance-X}
  d_X'=\min(d_X,d_X^{(0)}),\quad
  d_X^{(0)}=\min_{b \,:\, u_0b^T=1\,\wedge\, Q'b^T=0}  \wgt(b).
\end{equation}

The lower bounds constructed in the following two subsections both
rely on geometry in a bipartite (Tanner) graph associated with the
type-$Z$ generator matrix $H_Z=Q$.  Namely, given its row-set $U$
(check-nodes) and column-set $V$ (value-nodes), the Tanner graph has
the union $U\cup V$ as its vertex set, and an undirected edge
$(u,v)\in U\times V$ for each non-zero matrix element $Q_{uv}$.  On a
graph there is a natural notion of the distance between a pair of
nodes, the number of edges in the shortest path between them; a ball
$\Omega_{R}(u_0)$ of radius $R$ centered around $u_0$ is the set of
all vertices at distance $R$ or smaller from $u_0$.  Then, an erasable
region $A=\Omega_R(u_0)\cap V$ is chosen as a set of value nodes
within the radius $R$ from a check node $u_0\in U$, subject to the
condition that a row of the shortened matrix $Q_B$ contains $u_0$ in
its expansion over the rows of $Q$. 

The condition is not a trivial one, as it is actually equivalent to
region $A$ being erasable in the code $\css(P,Q')$ formed by the
original matrix $H_X\equiv P$ and the matrix $Q'$, the original matrix
$Q$ with the row $u_0$ (considered linearly independent) dropped, same
code as in Eqs.~(\ref{eq:row-distance-Z}) to
(\ref{eq:CSS-distance-X}).  As an equivalent but easier to check
condition, one may request that row $u_0[A]$ be a linear combination
of the rows of the punctured matrix $Q'[A]$ (remember that the support
of $u_0$ is a subset of $A$, while $u_0$ is linearly independent from
the rows of $Q'$).  In addition, we use a corresponding sufficient
condition as a part of lower $X$-distance bound in Statement
\ref{th:lower-dX-bound}, and formulate a related necessary condition
in terms of the topological entanglement entropy associated with the
defect $A$ in Sec.~\ref{sec:tee}.

\subsection{Code with locally linearly-independent generators}
\label{sec:dX-bound}

We need a condition to guarantee a lower bound on the weight of the
operator conjugate to the row $u_0$ removed from the matrix ${Q}_B$,
see Eq.~(\ref{eq:CSS-distance-X}).  Here, we will assume that the set
of $Z$-type stabilizer generators forming the rows of the matrix
$H_Z=Q$ be overcomplete.  That is, there be one or more linear
relations between the rows of $Q$, and that we start with a row $u_0$
which takes part in such a relation.

\begin{figure}[h]
  \centering
  \includegraphics[width=\textwidth]{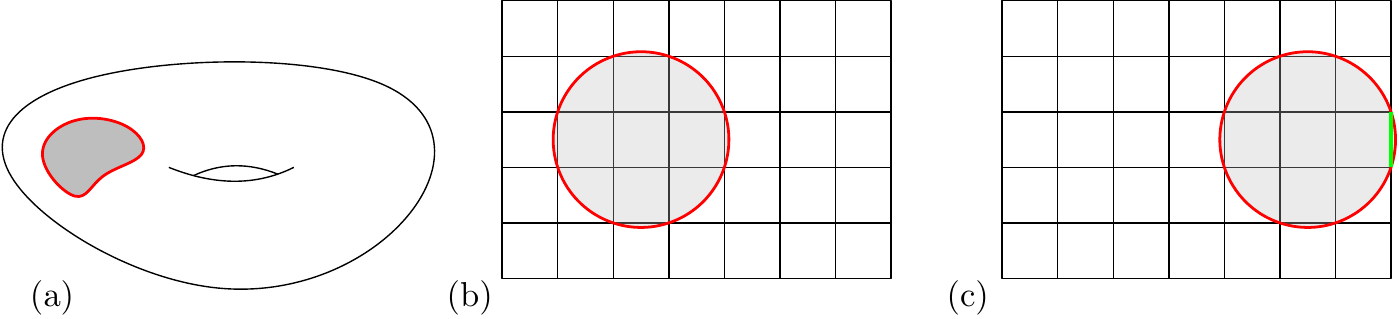}
  \caption{(a) Homologically trivial hole on a torus.  (b) Surface
    code with a smooth boundary.  Removing qubits (edges) inside of
    the circle we get a non-trivial defect.  (c) This circle contains
    a boundary edge with no neighboring plaquette; removing the
    corresponding edges we again get a trivial defect.}
  \label{fig:surf}
\end{figure}

In the case of the toric code (or any surface code on a locally planar
graph without boundaries), see Fig.~\ref{fig:surf}(a), the linear
relation is simply the statement that the sum of all rows of $H_Z$ be
zero (necessarily so since each column has weight two).  Such a
relation exists for any matrix with even column weights, e.g., qHPs
from $(\ell,m)$-regular binary codes with both $\ell$ and $m$ even.
Further, many such linear relations exist for CSS codes forming chain
complexes of length $3$ or more, e.g., the $D$-dimensional
hyperbolic\cite{Guth-Lubotzky-2014,Breuckmann-thesis-2017} and
higher-dimensional qHP codes\cite{Zeng-Pryadko-2018} with $D>2$.
\begin{statement}
  \label{th:lower-dX-bound}
  Given a CSS code $\css(P,Q)$ and a natural $R_1$, consider the
  bipartite Tanner graph associated with the matrix $Q$, and a ball
  $W=\Omega_{2R_1}(u_0)$ of radius $2R_1$ centered around the row
  $u_0\in U$.  Assume {\rm (\textbf{a})} that the row $u_0$ is involved in
  at least one linear relation with other rows of $Q$, and
  {\rm (\textbf{b})} there exists $R_2>R_1$ such that all rows within
  radius $2R_2$ from the center be linearly independent of each other.
  Let ${Q}_1$ denote a full-row-rank matrix obtained from $Q$ by
  removing some (linearly-dependent) rows outside $W$. Then weight of
  any $b\in \mathbb{F}_2^{\otimes n}$ such that the syndrome $Q_1 b^T$
  has the only non-zero bit at the check node $u_0$ satisfies
  $\wgt(b)\ge R_2$, and for the complement $B=V\setminus A$ of any
  region $A\subseteq W$, $\wgt(b[B])\ge R_2-R_1$.
\end{statement}
The punctured vector $b[B]$ is a representative of the new $X$-type
codeword in the defect code $\css(P[B],Q_B')$, where $Q'$ is obtained
from $Q_1$ by removing $u_0$; the constructed bound gives
$d_X^{(0)}\ge R_2-R_1$ for the distance in
Eq.~(\ref{eq:CSS-distance-X}).  

We also note that additional, linearly-dependent with $u_0$, rows in
$Q$ need not have bounded weight, as long as on the Tanner graph they
are located outside the ball $W_2$.  The corresponding requirement is
of course equivalent to any of the two conditions above the subsection
\ref{sec:dX-bound} title, with $A=W_2\cap V$.  In the case of a surface
code with smooth boundary, see Fig.~\ref{fig:surf} (b) and (c), the
extra row may be chosen as the product of all plaquette generators,
with the support along the actual boundary.  In such a case, the lower
distance bound in Statement \ref{th:lower-dX-bound} is saturated.

\subsection{Stabilizer group with an expansion}
Here we construct a simple lower bound on the $Z$-distance of the
defect, in essence, relying on the monotonicity of the distance $d_Z$
with respect to $X$-basis measurement of qubits in an erasable set,
see Statement \ref{th:CSS-decomposition}.  To make it non-trivial, we
assume that $Z$-type stabilizer generators of the original code
satisfy an expansion condition, namely, there exists an increasing
real-valued function $f$ such that a product $\Pi_m$ of any $m$
distinct generators has weight bounded by $f(m)$,
\begin{equation}
\wgt(\Pi_m)\ge f(m),\quad f(m+1)> f(m).\label{eq:expansion-condition}
\end{equation}
Such a \emph{global} condition on code generators guarantees that the
boundary condition is good for the defect we are trying to construct.
For example, in case of the toric code on an $L\times L$ square
lattice with periodic boundary conditions, there are $L^2$ plaquette
generators but only $L^2-1$ of them are independent.  Namely, the
product of all plaquette generators is an identity, so $m\to L^2-m$ is
a symmetry of the weight distribution.  Necessarily, the function $f$
in Eq.~(\ref{eq:expansion-condition}) has a trivial maximum,
$f(L^2)\le 0$.  Respectively, a single hole in Fig.~\ref{fig:surf}(a)
has a homologically trivial boundary---meaning that it can be pushed
out and eventually contracted to nothing by a sequence of
single-plaquette steps.  On the other hand, for a planar
smooth-boundary surface code configuration as in
Fig.~\ref{fig:surf}(b), one gets $f(m)$ scaling as a perimeter of $m$
plaquettes with a non-trivial maximum.

Generally, as one increases the set $A$ of removed qubits, there will
be rows in the shortened matrix $Q_B$ formed as linear combinations of
increasing numbers of rows of the original matrix $Q$.  The expansion
condition (\ref{eq:expansion-condition}) with $\max_m f(m)>0$
guarantees that the corresponding rows cannot be contracted to
nothing.  For example, when we remove a single qubit corresponding to
a weight-$\ell$ column of $Q$, if the corresponding adjacent rows all
have weights $w$ and do not overlap (in the case of a surface code
$\ell\le 2$ and $w$ is the number of sides in the corresponding
plaquette), the shortened matrix $Q_B$ necessarily has $\kappa=\ell-1$
rows of weight $2w-2$.  Assuming $f(2)=2w-2$, at any $w>2$ this is
already sufficient to guarantee the existence of a degenerate defect
code with $d_Z'>w$.

With larger defects, combinations of larger numbers of rows may become
necessary.  If so, the expansion condition
(\ref{eq:expansion-condition}) will also guarantee that codes with
$Z$-type distances (\ref{eq:CSS-distance-Z}) much greater than $w$ can
be constructed (assuming big enough original code distance $d_Z$).
\begin{statement}
  \label{th:lower-dZ-bound-new}
  Given a code $\css(P,Q)$ and a natural $R_1$, consider the bipartite
  Tanner graph associated with the matrix $Q$, and a ball
  $W=\Omega_{2R_1}(u_0)$ of radius $2R_1$ centered around the row
  $u_0\in U$.  Denote $Q'$ the matrix obtained by removing $u_0$ from
  $Q$.  Assume {\rm   ({\bf a})} that $A\equiv W\cap V$ is erasable in the
  code $\css(P,Q')$, and {\rm ({\bf b})} that the set of $Z$-generators
  defined by the rows of matrix $Q$ satisfies the expansion condition
  (\ref{eq:expansion-condition}) with $f(2)\ge 1$.  Then, weight of
  any linear combination of $u_0$ with rows of the matrix $Q'$
  supported on the complement $B\equiv V\setminus A$ satisfies
  $d_Z^{(0)}\ge f(R_1)$.
\end{statement}
Notice that the condition (\textbf{a}) here is the same as discussed
above the Section \ref{sec:dX-bound} title; the corresponding
sufficient condition is a part of Statement \ref{th:lower-dX-bound},
where any $R_2>R_1$ will do. 

\subsection{Defect codes with arbitrary large distances}
Notice that Statement \ref{th:lower-dX-bound} requires a linear
\emph{dependence} between generators of $Q$, while Statement
\ref{th:lower-dZ-bound-new} requires the expansion condition
(\ref{eq:expansion-condition}) with non-trivial $f$ which is stronger
than just linear \emph{independence}.  Nevertheless, these conditions
are not necessarily incompatible.  The condition in Statement
\ref{th:lower-dX-bound} only needs to be satisfied for some parent
code.  For example, in the case of a toric code, \emph{two} distinct
holes are needed in order to create a defect code with distance
$d'=\min(d_Z',d_X')>4$.  Here a linear dependence between plaquette
operators can be found in the parent toric code, one hole is needed to
satisfy the conditions of Statement \ref{th:lower-dZ-bound-new}, while the
other one is the actual erasable set in
Construction~\ref{constr:defect}.

Generally, suppose we have a parent CSS code $\css(H_X,H_Z)$ with
bounded-weight generators, sufficiently large distance, and matrix
$H_Z$ with even-weight columns so that the sum of all rows be
zero. 
Such a pair of matrices satisfies conditions of Statement
\ref{th:lower-dX-bound} but not of Statement
\ref{th:lower-dZ-bound-new}.  Similar to the toric code, where one
needs two holes to create a single-qubit defect, here we also may need
to take an erasable set $A$ formed by two or more disjoint erasable
defects, e.g., balls as in Statement \ref{th:lower-dZ-bound-new}; that
the set $A$ be erasable can be guaranteed by the union Lemma (Lemma 2
in Ref.~\onlinecite{Bravyi-Terhal-2009}).  Then, one (or more if
needed) balls can be used to ensure the existence of the function $f$
in Eq.~(\ref{eq:expansion-condition}) with sufficiently large
$\max_m f(m)$, while the qubits in the last remaining ball would be
used as the erasable set.

Explicitly, as a parent code family, one can use, e.g., qHP
codes\cite{Tillich-Zemor-2009} created from random matrices with
even-valued row and column weights.  For example, $(4,6)$-regular
random matrices would do well, leading to qHP codes with
asymptotically finite rates, whose CSS generator matrices have column
weights $4$ and $6$, regular row weights $w= 10$, and
${\cal O}(n^{1/2})$ linear relations between the rows of generator
matrices with number of non-zero coefficients in each linear relation
scaling linearly with block length $n$ of the resulting code.  The
distance of such parent codes grows as ${\cal O}(n^{1/2})$; this is
sufficient to ensure that for any $d_0>0$ one can choose $n$ large
enough so that sufficiently large erasable balls exist to guarantee
the existence of defect codes with $d'\equiv \max(d_X',d_Z')\ge d_0$.

\section{Relation with topological entanglement entropy}
\label{sec:tee}

There exists a suggestive parallel between the structure of a
large-distance qubit-carrying defect we discussed, and (generalized)
topological entanglement entropy (TEE) which can be associated with
such a defect\cite{Kitaev-Preskill-2006,Levin-Wen-2006}.  The latter
may be defined in terms of the usual entanglement entropy (EE), which
characterizes what happens when some of the qubits carrying a
normalized quantum state $\ket\psi\in \mathcal{H}_2^{\otimes n}$ are
erased (traced over).  Namely, if the set of qubits is decomposed into
$A$ and its complement $B=V\setminus A$, one considers the binary von
Neumann entropy $\Upsilon(A;B)\equiv -\tr_B\rho_B\log_2\rho_B$, where
the density matrix $\rho_B=\tr_A\ket\psi\!\bra\psi$ is obtained by
tracing over the qubits in $A$.  The definition is actually symmetric
with respect to interchanging $A$ and $B$,
$\Upsilon(A;B)=\Upsilon(B;A)$.

The entanglement entropy has a particularly simple form when
$\ket\psi$ is a stabilizer
state\cite{Fattal-Cubitt-Yamamoto-Bravyi-Chuang-2004}.  Such a state
is just a stabilizer code encoding no qubits, so that its dimension is
$2^0=1$.  With $n=|V|$ total qubits, this requires a stabilizer group
with $n$ independent generators.  According to Fattal et
al.\cite{Fattal-Cubitt-Yamamoto-Bravyi-Chuang-2004}, the EE of any
stabilizer state $\ket\psi\in\mathcal{Q}$ is uniquely determined by
the decomposition of the stabilizer group
$\mathcal{S}=\mathcal{S}_A\times \mathcal{S}_B\times
\mathcal{S}_{AB}$, where non-trivial elements of subgroups
$\mathcal{S}_A$ and $\mathcal{S}_B$ are supported only on $A$ and only
on $B$, respectively, and those of $\mathcal{S}_{AB}$ are necessarily
split between $A$ and $B$.  Namely, $\rank\mathcal{S}_{AB}=2p$ is
always even, and it is this $p$ (or, equivalently, the number of EPR
pairs split between $A$ and $B$) that determines the entanglement
entropy,
\begin{equation}
\Upsilon(A;B)=p\equiv {1\over 2}\rank \mathcal{S}_{AB}.\label{eq:ee}
\end{equation}
Given a stabilizer code $\mathcal{Q}$ with parameters $[[n,k,d]]$ and
a stabilizer group $\mathcal{S}$ of rank $n-k$, a stabilizer state
$\ket\psi\in\mathcal{Q}$ can be formed by adding any $k$ mutually
commuting logical Pauli operators to the stabilizer group.  Then, if
set $A\subset V$ is erasable, according to the cleaning
lemma\cite{Bravyi-Poulin-Terhal-2010}, we can select all logical
operators with the support in $B=V\setminus A$.  With the logical
operators in ${\cal S}_B$, both ${\cal S}_A$ and ${\cal S}_{AB}$ are
subgroups of the stabilizer group ${\cal S}$ of our original code, and
the entanglement entropy is given by the same Eq.~(\ref{eq:ee}).  The
same quantity $p$ can also be expressed in terms of the punctured
stabilizer group ${\cal S}[B]={\cal S}_B\times {\cal S}_{AB}[B]$
(gauge group of the subsystem erasure code), written as a product of
its center, the (shortened) stabilizer group ${\cal S}_B$, and $p$
pairs of canonically conjugated ``gauge'' qubits which generate the
(punctured) subgroup ${\cal S}_{AB}[B]$.

In the case of a CSS code, such a decomposition exists for both
$X$-type and $Z$-type subgroups of the stabilizer, e.g.,
${\cal S}^{(Z)}={\cal S}^{(Z)}_A\times {\cal S}^{(Z)}_B\times {\cal
  S}^{(Z)}_{AB}$, with
$\rank {\cal S}^{(Z)}_{AB}=\rank {\cal S}^{(X)}_{AB}=p$.  These $p$
independent generators are obtained from the rows of the original
generator matrices that are split between $A$ and $B$.  In a
weight-limited LDPC code, the total number of such rows, e.g., in
$H_Z$, can be called the perimeter $L(A;B)$ of the cut.  However, the
number of generators of $\mathcal{S}_{AB}^{(Z)}$ can actually be
smaller than $L(A;B)$ since some linear combination(s) of the generators
split between $A$ and $B$ combined with other generators may form an
element of ${\cal S}_A$ or ${\cal S}_B$.  Thus, we can write EE as
\begin{equation}
\Upsilon(A;B)=L-\gamma,\label{eq:ee-expansion}
\end{equation}
with $L=L(A;B)$ the perimeter of the cut and some integer
$\gamma\equiv \gamma(A;B)\ge0$.  While this expression strongly
resembles the Kitaev-Preskill definition of TEE
\cite{Kitaev-Preskill-2006,Levin-Wen-2006}, for now $\gamma$ is just a
parameter associated with the particular cut.

Let us now consider weights of generators of ${\cal S}_B^{(Z)}$.
These correspond to the rows of $Q_B$.  Clearly, each row of the
original matrix $Q$ may be supported in $A$, or in $B$, or be split
between the two sets.  Rows already supported in $B$ can be moved
directly to $Q_B$ and preserve their original weights.  Thus no more
than $\gamma\ge0$ generators of ${\cal S}_B^{(Z)}$ may need to have
larger weights.  Necessarily, if we want to construct a defect forming
a degenerate code with the distance $d'>w$, the additional number of
qubits is bounded by
\begin{equation}
  \kappa\equiv k'-k\le \gamma.
  \label{eq:tee-bound}
\end{equation}
Thus, with $\gamma=0$, the defect cannot support a degenerate code
with $\kappa>0$.  However, whether or not
a particular defect does, in fact, support $\kappa>0$,
also depends on the global structure of the code, e.g., the boundary
conditions.

Now, let us imagine that we have a defect code with a sufficiently
large distance $d$. Then, such a defect is also \emph{stable} to small
deformations, e.g., when $B$ is changed to some $B'$ as a result of up
to $M<d$ steps, where at each step a single position is added or
removed from the set.  
That is, our defect code retains the same
number $\kappa$ of additional qubits when we change the set $B$ to a
set $B'$, $|B\triangle B'|\le M$, where
$B\triangle B'=(B\setminus B')\cup(B'\setminus B)$ is the symmetric
set difference.  For deformations such that $M+w<d$, the inequality
$\gamma\ge \kappa$ must be satisfied in the course of deformations.

Now, TEE is normally considered a property of ground-state wave
function of some many-body Hamiltonian, while our focus was on quantum
LDPC codes with bounded-weight but not necessarily local generators.
Different terms in a Hamiltonian can be viewed as generators of the
code.  However, in the absence of locality, why would we care about
weights of terms in a quantum spin Hamiltonian?

 In a physical system, multi-qubit Pauli operators may appear as terms
 in an $n$-spin quantum Hamiltonian, e.g.,
\begin{equation}
  H_0=- A\sum_{a}  P_a-B\sum_b Q_b,  \label{eq:hamiltonian}
\end{equation}
where and $A>0$ and $B>0$ are the coupling constants, and, to connect
with our discussion of CSS codes, $P_a$ and $Q_b$ could be Pauli
operators of $X$- and $Z$-type, respectively, specified by rows of the
binary matrices $P$ and $Q$.  Then, if all terms in the Hamiltonian
commute, i.e., $PQ^T=0$, the ground state space of $H_0$ is exactly
the code with the stabilizer group generated by these operators.

Any simple spin Hamiltonian (\ref{eq:hamiltonian}) is usually
just the leading-order approximation to a real problem.  Even at zero
temperature, additional interaction terms are virtually always
present.  Such terms may break the degeneracy of the ground state of
the Hamiltonian $H_0$.  The effect is weak if the code has a large
distance, while perturbations be small and local.  The standard
example is the effect of an external magnetic field
$\mathbf{h}=(h_x,h_y,h_z)$, which can be introduced as an additional
perturbation Hamiltonian
\begin{equation}
  \label{eq:perturbation}
  H_1=-{1\over2}\sum_i (h_x X_i+h_y Y_i+h_z  Z_i).
\end{equation}
For a code with distance $d$, only a Pauli operator of weight $d$ or
larger may act within the code.  Respectively, assuming the magnetic
field small, degenerate perturbation theory gives the ground state
subspace energy splitting scaling as ${\cal O}( h^d)$, where
$h=|\mathbf{h}|$ is the field magnitude.

However, the code distance $d$ gives only a part of the story.
Large-weight operators appearing in $H_0$ make the ground-state order
particularly susceptible to local perturbations such as the magnetic
field.  In this case the relevant scale for the magnetic field is
$Wh\sim \max(A,B)$, that is, the effect of the magnetic field may be
magnified by the operator weight $W$.  Indeed, if we start with the
spin-polarized ground state of $H_1$, a weight-$W$ Pauli operator will
generically flip $W$ spins, producing a state with the energy
increased by ${\cal O}(Wh)$.  The effect of such a perturbation will
be small as long as the corresponding coefficient, $A$ or $B$ in
Eq.~(\ref{eq:hamiltonian}), remains small compared to $Wh$.
Thus, with $W$ large, the ground state of the spin Hamiltonian $H_0$ gets
destroyed already with very small  $h\sim \max(A,B)/W$. 
The same estimate can be also obtained with the help of an
  exact operator map similar to that used by Trebst et
  al.\cite{Trebst-tension-2007}.

\section{Conclusions}
To summarize, we discussed a general approach to adding logical qubits
to an existing quantum stabilizer code, with the focus on quantum LDPC
codes with weight-limited stabilizer generators.  In short, a
stabilizer generator needs to be promoted to a logical operator, which
puts a bound on the distance of the obtained code in terms of the
generator weight $w$.  As in a surface code, a degenerate code
can be obtained by removing some qubits in an erasable set, and
gauge-fixing the resulting subsystem code in such a way as to ensure
that stabilizer generators of sufficiently large weight be created.
We also constructed some lower bounds on the distance of thus obtained
defect codes which show that construction can in principle be used to
obtain highly degenerate codes with distances much larger than $w$.

An interesting observation is a relation between the ability of a
particular defect (erasable set of qubits) to support an additional
logical qubit in a degenerate code, and a quantity analogous to TEE,
$\gamma$.  A degenerate defect code can be only created with
$\gamma>0$.  Further, when a defect code has a large distance $d'$, a
lower bound on $\gamma>0$ is maintained in the course of deformations,
not unlike for the conventionally defined TEE.

 Many open problems remain.  First, our lower distance bounds
are constructed by analogy with surface codes.  In particular, the
lower bound in Statement \ref{th:lower-dX-bound} applies only for a
single qubit.  In addition, we do not have good lower distance bounds
for defects in non-CSS codes.  

Second, the notion of generalized TEE $\gamma$ in
Eq.~(\ref{eq:ee-expansion}) needs to be cleaned up.  Here we are
working with lattice systems, not necessarily local, and the usual
expansions in term of $1/L$ do not necessarily help.  Further, as
defined, $\gamma$ certainly depends of the chosen set of generators.
Redundant sets of small-weight generators imply the existence of
higher homologies, as in higher-dimensional toric codes; it would be
nice to be able to interpret values of $\gamma$, as, e.g., was done by
Grover et al.\ in a field theory
setting\cite{Grover-Turner-Vishwanath-2011}.

Third, if we start with a finite-rate family of codes, are there
defects of size $|A|$ with $\gamma={\cal O}(|A|)$?  Coming back to
defect codes, it appears that a typical defect with large $\gamma$
would generically lead to an entire spectrum of operator weights in
the generators of ${\cal S}_B$.  Is there a situation when there is a
large gap in this weight distribution, as in the surface codes with
$\gamma=1$, where only one high-weight operator may exist?

\medskip\noindent\textbf{Acknowledgment:} This work was supported in part by the NSF
Division of Physics via grant No.\ 1820939.

\begin{appendix}
\section{All the proofs}
\label{sec:proofs}
\begin{proof}[Proof of Statement \ref{th:CSS-decomposition}] The
  number of encoded qubits follows from the identity
  (\ref{eq:puncture-shortening-rank}).  Namely, the exact dual $Q^*$
  of the matrix $Q$ can be obtained from $P$ by adding $k$ rows
  corresponding to inequivalent codewords
  $b\in{\cal C}_Q^\perp\setminus {\cal C}_P$.  According to the
  cleaning lemma \cite{Bravyi-Poulin-Terhal-2010}, these can be chosen
  with the support outside of an erasable set $A$.  Dropping these $k$
  rows from $Q^*[B]$ recovers the punctured matrix $P[B]$ with the
  correct rank to ensure $k'=k$.  The distance inequalities are
  obtained from Eqs.~(\ref{eq:CSS-distance}) by considering removal of
  a single qubit at a time.
\end{proof}

\begin{proof}[Proof of Statement \ref{th:lower-dX-bound}]
  Indeed, since $Q'$ differs from $Q$ only by some rows outside the
  ball $W_2\equiv \Omega_{2R_2}(u_0)$ which are linearly dependent
  with $u_0$, the corresponding full-matrix syndrome $Q b^T$ must have
  non-zero bits outside $W_2\cap U$.  With the exception of $u_0$, any
  row in $W_2\cap U$ must be incident on an even number of set bits in
  $b$, and there must be a continuous path on the graph from $u_0$ to
  outside $W_2$ formed by pairs of set bits in $b$ (otherwise $b$
  could be separated into a pair of vectors with non-overlapping
  supports, $b=b_1+b_2$, such that $Q b_1^T=0$ outside $W_2$ and
  $Q b_2^T=0$ inside $W_2$, which would contradict the assumptions).
  This guarantees that at any odd distance from $u_0$ (up to
  $2R_2-1$), $b$ has at least one set bit, which recovers the two
  lower bounds.
\end{proof}
\begin{proof}[Proof of Statement \ref{th:lower-dZ-bound-new}]
  The inequality follows from the locality of row operations on the
  Tanner graph: a non-zero bit $v$ in some vector
  $c\in \mathbb{F}_2^{\otimes n}$ can only be removed by adding a row
  $u\in U$ neighboring with $v$.  The lower bound on $f$ equivalent to
  linear independence of rows of $Q$ guarantees that rows of $Q'$ be
  linearly independent from $u_0$, thus weight must remain non-zero at
  every step.    The condition (\textbf{a}) guarantees the existence
of a linear combination in question.

\end{proof}

\end{appendix}

\bibliographystyle{ws-rv-van}
\bibliography{lpp,qc_all,more_qc,spin,percol}

\end{document}